\documentclass[a4paper,twoside,10pt]{scrreprt}

\usepackage[utf8]{inputenc}		
\usepackage[ngerman]{babel}	
\usepackage{babelbib}			
\usepackage{float}
\usepackage{pifont,textcomp}		
\usepackage{paralist}			
\usepackage{graphics,float}
\usepackage{graphicx}
\usepackage{tabularx}
\usepackage{listings}			
\usepackage{xcolor}
\usepackage{keystroke}			
\usepackage{rotating}			
\usepackage{verbatim}
\usepackage{footmisc,booktabs}	
\usepackage[acronym, toc, section, nonumberlist, nopostdot]{glossaries}
\usepackage[hyphens]{url}
\usepackage{lscape}
\usepackage{multirow}
\usepackage{multicol}
\usepackage{subfigure}
\usepackage[11pt]{moresize}
\usepackage{bytefield}
\usepackage{todonotes}
\usepackage[absolute]{textpos}
\usepackage{fancybox}
\usepackage{framed}
\usepackage{morewrites} 
\usepackage{amsmath}
\usepackage{colortbl}
\usepackage{geometry}
\usepackage{fancyhdr}
\usepackage{rotating}
\usepackage{afterpage}
\usepackage{pdflscape}
\usepackage[pagebackref=false]{hyperref}
\usepackage{pdflscape} 
\usepackage{array}
\usepackage{longtable}
\usepackage{multicol}
\usepackage[font=scriptsize,labelformat=empty,textfont=it,up]{caption}
\usepackage[export]{adjustbox}
\usepackage{parnotes}
\usepackage{marvosym}

\newcolumntype{L}[1]{>{\raggedright\let\newline\\\arraybackslash\hspace{0pt}}m{#1}}
\newcolumntype{C}[1]{>{\centering\let\newline\\\arraybackslash\hspace{0pt}}m{#1}}
\newcolumntype{R}[1]{>{\raggedleft\let\newline\\\arraybackslash\hspace{0pt}}m{#1}}
\newcolumntype{x}{>{\centering\let\newline\\\arraybackslash\hspace{0pt}}X}

\hypersetup{backref=false,  
  colorlinks=true,  
  linkcolor=black,    
  citecolor=black,    
  filecolor=black,     
  urlcolor=black,   
  pdfauthor=Julian Schütte, 
  pdfproducer=PDFLatex
}

\definecolor{fraunhoferblue}{rgb}{.2,.47,.72}
\definecolor{fraunhofergray}{rgb}{.8,.8,.8}
\definecolor{fraunhofergreen}{rgb}{0.09,.61,.49}
\definecolor{fraunhofergreengray}{rgb}{0.03,.20,.18}
\definecolor{fraunhofergreenlight}{rgb}{0.23, 0.92, 0.75}

\fancypagestyle{plain}{%
\fancyhead{}
\fancyfoot{}
\renewcommand{\headrulewidth}{0.4pt}
\renewcommand{\footrulewidth}{0.4pt}
\fancyhead[RO, LE] {\nouppercase{\leftmark{}}\vspace*{-.35cm}}
\fancyfoot[RO, LE] {\thepage}
\fancyfoot[LO, RE] {Fraunhofer AISEC}
}
\renewcommand{\headrule}{\hbox to\headwidth{\color{fraunhofergreen}\leaders\hrule height \headrulewidth\hfill}}
\renewcommand{\footrule}{\hbox to\headwidth{\color{fraunhofergreen}\leaders\hrule height \footrulewidth\hfill}}
\usepackage{eso-pic}

\usepackage{draftwatermark}
\SetWatermarkAngle{45}
\SetWatermarkLightness{0.8}
\SetWatermarkFontSize{5cm}
\SetWatermarkScale{1}
\SetWatermarkText{} 

\renewcommand{\arraystretch}{1.2}

\lstdefinestyle{tcblatex}{
  moredelim=**[is][\color{red}]{@@}{@@},
  moredelim=**[is][\rule{.9\textwidth}{0.3pt}]{!!}{!!},
}

\usepackage[most]{tcolorbox}

\newtcblisting[auto counter]{nicelisting}[3][]{%
  listing options={style=tcblatex,#1},hbox,listing only,
  enhanced,arc=4pt,outer arc=4pt,top=1mm,bottom=1mm,left=1mm,right=1mm,
  boxrule=0.6pt,drop fuzzy shadow,fontupper=\tiny,fontlower=\tiny,before=\begin{center},after=\end{center},
  title=Listing \thetcbcounter: #2, label={#3}}

\newtcbinputlisting{\niceinputlisting}[2][]{%
 listing options={style=tcblatex,#1},hbox,listing only,
  enhanced,arc=4pt,outer arc=4pt,top=1mm,bottom=1mm,left=1mm,right=1mm,
  boxrule=0.6pt,drop fuzzy shadow,listing file=#2%
}

\usepackage{xifthen}
\newcounter{ProtStepCounter}

\newcommand{\step}[4][]
{
  \stepcounter{ProtStepCounter}\arabic{ProtStepCounter}: &
  #2 $\to$ #3 :&
  #4 \\
  \ifthenelse{\isempty{#1}}{}{ & &(\footnotesize\textit{#1})\\}
}
\newcommand{\define}[2][]
{
   &
  \ifthenelse{\isempty{#1}}{}{#1:}

   &
  #2 \\
}

\usefont{T1}{pfr}{l}{n}\selectfont
\setkomafont{chapter}{\usefont{T1}{pfr}{m}{n}\selectfont\color{black}\Large}
\addtokomafont{section}{\usefont{T1}{pfr}{m}{n}\selectfont\color{black}\large}
\addtokomafont{subsection}{\usefont{T1}{pfr}{m}{n}\selectfont\color{black}\bfseries}

\addto{\captionsngerman}{ }
\addto{\captionsngerman}{ }

\makeatletter

\newenvironment{Figure}
  {\par\medskip\noindent\minipage{\linewidth}}
  {\endminipage\par\medskip}

\newcommand\blankpage{%
    \null
    \thispagestyle{empty}%
    \addtocounter{page}{-1}%
    \newpage}

\let\olditemize=\itemize
\def\itemize{
        \olditemize
        \setlength{\itemsep}{-0.8ex}
        }

\title{Der Trusted Connector im Industrial Data Space}
\subtitle{Datensouveränität im Internet der Dinge}
\publishers{Fraunhofer AISEC}
\author{Julian Schütte, Gerd Brost, Sascha Wessel}
\date{\today}


 \makeindex
 \makeglossaries
\begin{document}

\maketitle



\chapter{Sicherheit im Industriellen Internet der Dinge}
Die Digitalisierung durchdringt alle Branchen und führt zu tiefgreifenden Veränderungen von Geschäftsmodellen und technischen Infrastrukturen. Insbesondere in Logistik und Produktion bietet die Vernetzung von Geräten und der Austausch von Daten über Unternehmensgrenzen hinweg die Chance, Abläufe zu beschleunigen, Kosten zu reduzieren und Medienbrüche zu vermeiden. Das hierbei entstehende \emph{Industrielle Internet der Dinge} (IIoT) bringt jedoch völlig neue Anforderungen an den Datenschutz und die Sicherheit der technischen Infrastrukturen mit sich. Diese Anforderungen adressiert die Initiative des \glqq Industrial Data Space\grqq{} \cite{IDS2016} mit dem Trusted Connector.

\vspace*{2em}
\begin{multicols}{2}

\paragraph{Datenschutz}
Im industriellen Internet der Dinge (IIoT) werden Daten über
Unternehmensgrenzen hinweg ausgetauscht. Unternehmensgrenzen sind dabei auch
meist Vertrauensgrenzen, aus denen Daten herausgegeben werden. Hierzu zählen
auch und insbesondere sensible Daten, aus denen sich Rückschlüsse über
Geschäftsgeheimnisse ziehen lassen, wie beispielsweise Angaben über verfügbare
Produktionskapazitäten, Lieferketten oder Wartungsdaten von Sensoren. Aber auch
persönliche Daten wie Bewegungsprofile oder Gesundheitsdaten werden
herangezogen und für Analysezwecke verarbeitet.

Zeitgleich sehen sich Unternehmen einer verschärften Datenschutzgesetzgebung gegenüber, nach der die Auskunft über den Verbleib persönlicher Daten und ihre Löschung auf Verlangen des Benutzers jederzeit möglich sein muss.

Um diesen Anforderungen gerecht zu werden, müssen Architekturen für das
Internet der Dinge die Herkunft von Daten verfolgen und es Benutzern
ermöglichen, nicht nur den Zugriff, sondern auch die Art und Weise der
Verarbeitung von Daten zu kontrollieren.

\paragraph{Sicherheit}

Wie die jüngste Vergangenheit gezeigt hat, sind eingebettete Systeme im Internet der Dinge häufig Einfallstor für Schadsoftware und Eintrittspunkt für gezielte Angriffe auf die interne Infrastruktur. Veraltete Software, das Fehlen von sicheren Update-Mechanismen sowie die Langlebigkeit solcher Geräte lassen sie zu einem attraktiven Ziel für Angreifern werden.

Verwundbarkeiten lassen sich niemals völlig ausschließen. In heterogenen Infrastrukturen kommt jedoch die Herausforderung hinzu, dass das Sicherheitsniveau eines Kommunikationspartners initial vollkommen unbekannt ist und zunächst ein gegenseitiges Vertrauen etabliert werden muss. Hierzu ist es erforderlich, dass sich Kommunikationspartner ihr technisches Sicherheitsniveau automatisch nachweisen, so dass auf dieser Basis verlässlich beurteilt werden kann, ob und welche Daten mit dem jeweiligen Kommunikationspartner geteilt werden dürfen.

Infrastrukturen für das industrielle Internet der Dinge müssen daher einerseits Geräte unterschiedlicher Sicherheitslevel zulassen, gleichzeitig jedoch Mechanismen anbieten, über die Diensteanbieter die Verwendung ihrer Daten auf weniger sicheren Plattformen zuverlässig einschränken können.

\paragraph{Vertrauen}

Der Austausch von sensiblen Daten erfordert Vertrauen in die Teilnehmer, aber
auch in die technische Infrastruktur selbst. Die Tatsache, dass Low-Cost-Geräte
wie Kameras oder Sensoren kompromittiert und für Angriffe auf kritische
IT-Infrastrukturen genutzt werden können, bedroht nicht nur die Sicherheit des
Gesamtsystems. Sie wirft auch die Frage auf, inwieweit Daten, die von solchen
Geräten stammen, vertraut werden kann und ob sie zuverlässig genug sind, um auf
ihrer Basis automatische Entscheidungen zu treffen. Datenaustausch über
Unternehmensgrenzen hinweg bedeutet meist auch Datenaustausch über die eigenen
Vertrauensgrenzen hinweg.
Es sind daher Maßnahmen erforderlich, durch die das Vertrauen in die Sicherheit der Komponenten im industriellen Internet der Dinge nicht nur gesteigert, sondern nachweisbar quantifizierbar wird. Erst durch eine automatische und zuverlässige Attestierung des Sicherheitsniveaus einer Komponente können Diensteanbieter entscheiden, ob sie Daten für diese Komponente freigeben.
Der Industrial Data Space ist eine Initiative der Industrie und der Fraunhofer-
Gesellschaft zum Aufbau einer Infrastruktur für den sicheren Austausch von
IIoT-Daten. Der Trusted Connector (\glqq Konnektor\grqq{}) bietet hierbei einen
Software-Stack für vertrauenswürdige Edge-Gateways, die sowohl im Industrial
Data Space, aber auch in anderen IIoT-Infrastrukturen eingesetzt werden können.
Je nach Ausprägung kann der Trusted Connector-Stack zum Aufbau von sicheren
eingebetteten Edge-Devices mit oder ohne Hardware-Vertrauensanker oder als
ressourcenstarker Cloud-Dienst für das Hosting anspruchsvoller
Datenanalyse-Anwendungen eingesetzt werden.
Er wurde im Hinblick auf typische Sicherheitsanforderungen des IIoT entworfen und bietet Lösungen für das Identitätsmanagement von Teilnehmern und Gateways, geschützte Ausführungsumgebungen für Anwendungen, sowie eine benutzerzentrische Kontrolle über den Zugriff auf Daten und die Art und Weise ihrer Nutzung.

\end{multicols}


\chapter{Sicherheitseigenschaften des Trusted Connector}

Der Trusted Connector spezifiziert eine Softwareplattform für
vertrauenswürdige, sichere IIoT-Gateways. Diese sind
die Schnittstelle zwischen Datenquellen wie physischen Sensoren und
der externen Datenaustauschplattform. Während letztere in vielen Fällen durch eine
zentrale Cloud-Anwendung realisiert wird, legt der Industrial Data Space seinen Schwerpunkt auf die Datesouveränität und vermeidet eine zentrale Datenhaltung durch eine zwangsweise vertrauenswürdige Stelle.
Statt dessen werden Daten lokal im Connector vorgehalten und über direkte Verbindungen
zwischen Datenanbieter und -nutzer ausgetauscht. Neben der eigentlichen Vermittlung von
Daten sind IIoT-Gateways für die Vorverarbeitung (Aggregation, Filterung), das
Message-Routing, die Bereitstellung von Diensten in der IIoT-Infrastruktur und
die Herstellung sicherer Konnektivität verantwortlich. Gleichzeitig müssen
Konnektoren eine große Bandbreite von Deployment-Varianten abdecken: Die Referenzimplementierung des Trusted Connector ist daher nicht an eine bestimmte Hardware gebunden, sondern kann
aktuell bereits auf ARM, x86 und PowerPC-Plattformen aufgesetzt werden.

Anforderungen an sichere eingebettete IIoT-Gateways wurden u.a. durch Microsoft definiert\footnote{Galen Hunt and George Letey and Edmund B. Nightingale. The Seven Properties of Highly Secure Devices,Microsoft Research NExT Operating Systems Technologies Group, März 2017}
Die folgende Tabelle zeigt, wie diese Anforderungen durch den Trusted Connector erfüllt werden. Tatsächlich geht der Trusted Connector jedoch über diese Anforderungen hinaus, da er verschiedene Hardware-Plattformen und damit auch verschiedene Sicherheitsstufen zulässt, die dem jeweiligen Kommunikationspartner kommuniziert und nachgewiesen werden können.
\vspace*{.5cm}

{\footnotesize
\begin{tabular}{l!{\color{fraunhofergreen}\vrule}p{11.5cm}}
\arrayrulecolor{fraunhofergreen}\hline
\textbf{Anforderung} & \textbf{Motivation/Umsetzung} \\
\hline
 \multirow{2}{*}{Hardware-based Root of Trust} & \emph{Geräte benötigen eine nicht-fälschbare und nicht-übertragbare eindeutige Identität. Schlüsselmaterial wird durch Hardware-Vetrauensanker geschützt.}\\
  \cline{2-2} & SCDaemon-Komponente abstrahiert Zugriff auf Schlüsselmaterial und Zertifikate von zugrundeligender Software/Hardware. \\ 
  \hline
  \multirow{2}{*}{Small Trusted Computing Base} & \emph{Funktionen sollten auf einer möglichst kleinen vertrauenswürdigen Basis aufbauen.}\\
  \cline{2-2} & Vertrauenswürdige Basis sind Kernel \& Container Management Layer. \\
  \hline
  \multirow{2}{*}{Defense in Depth} & \emph{Ergänzende Sicherheitsfunktionalitäten in mehreren Schichten zur Abwehr von Angriffen.}\\
  \cline{2-2} & Whitelist-basierte Integritätsschutzmaßnahmen (Secure Boot, Container-Integritätsverifikation), Remote-Integritätsnachweis, Least-Privilege-based Isolation.\\
  \hline
  \multirow{2}{*}{Compartmentalization} & \emph{Funktionen dürfen sich nicht gegenseitig beeinflussen. Ein Fehler in einer Funktion darf keine Auswirkungen auf andere Funktionen haben.}\\
  \cline{2-2} & Isolation von Apps in Linux-Containern, Least-Privilege-based Isolation mittels Linux Security Module (LSM).\\
  \hline
  \multirow{2}{*}{Certificate-based Authentication} & \emph{Geräte müssen sich sicher und nicht-interaktiv -- also ohne Eingabe von Passwörtern -- authentisieren können.}\\
  \cline{2-2} & Automatisch ausgestellt ACME-Zertifikate für vertrauliche Kommunikation, Nicht-interaktiver Nachweis von Identitätsattributen mittels OAuth2.0 und Attribute Provider.\\
  \hline
  \multirow{2}{*}{Renewable Security} & \emph{Fehlerhafte Komponenten müssen im Betrieb aktualisiert werden können.}\\
  \cline{2-2} & Remote-Aktualisierung von Container-Images und Kernel  \\
  \hline
  \multirow{2}{*}{Reporting} & \emph{Fehler und sicherheitskritische Ereignisse müssen nicht-abstreitbar aufgezeichnet werden.}\\
  \cline{2-2} & Eventbasiertes Audit-Log mit Integritätsnachweis. \\
  \hline
\end{tabular}}
\vspace*{.5cm}

\begin{multicols}{2}

\paragraph{Kommunikation}

Konnektoren etablieren einen sicheren Kommunikationskanal zwischen Endpunkten.
Neben Verschlüsselung und Schutz der Datenintegrität muss im Rahmen der
Kommunikation das Sicherheitsniveau des Gegenübers nachgewiesen werden. Je nach
verfügbarer Hardware-Plattform wird ein Trusted Platform Module (TPM) für eine
Remote-Attestation der Plattform eingesetzt. Bietet die jeweilige
Plattform keinen solchen Hardware-Vertrauensanker, so kann der Trusted Connector
nichtsdestotrotz mit einem Software-basierten TPM verwendet werden. In diesem Fall kann sein Kommunikationpartner das ggf. schwächere Sicherheitsniveau der Plattform im Rahmen der Remote Attestation feststellen und auf dieser Grundlage entscheiden, ob er Daten für diese Plattform freigibt.
Zusammenfassend gilt:
\begin{itemize}
  \item Kommunikation zwischen Konnektoren ist integer, authentisch und vertraulich.
  \item Konnektoren weisen sich gegenseitig ihren Sicherheitsstatus nach.
  \item Datenaustausch kann auf vertrauenswürdige sichere Konnektoren begrenzt werden.
\end{itemize}

\paragraph{Datensouveränität}

Unter dem Begriff \glqq Datensouveränität\grqq{} wird verstanden, dass Datenanbieter bestimmen können, wer ihre Daten erhält, wie sie verarbeitet werden dürfen und an welchen Zweck und an welche Auflagen die Nutzung gebunden sein soll. Diese Entscheidung muss anhand von Richtlinien definiert werden können, die für den Benutzer verständlich und für Auditoren nachprüfbar sind. Sie müssen auf authentischen Informationen über die Identität und das Sicherheitsniveau des Consumer-Connectors basieren.
Im Bezug auf die Datansouveränität zeichnet sich der Konnektors im Wesentlichen duch folgende Attribute aus: 
\begin{itemize}
  \item Die Bereitstellung von Daten kann abhängig vom Sicherheitsniveau des Datenkonsumenten erfolgen.
  \item Die Bereitstellung von Daten kann jedoch auch an Auflagen bzgl. ihrer Verwendung gebunden werden.
  \item Unzulässige Arten der Datenverarbeitungen und Datenflüsse können unterbunden werden.
  \item Die Einhaltung von zulässigen Datenflüssen ist nachweisbar und auditierbar.
\end{itemize}

\paragraph{Anwendungssicherheit}

Konnektoren dienen als Ausführungsplattform für Apps, durch die Daten
innerhalb des Konnektors verarbeitet und für andere Konnektoren bereitgestellt
werden können. Die Herkunft und Integrität von Apps muss während der Installation geprüft werden, um Manipulationen von Apps oder das Einschleusen von Schadsoftware zu verhindern. Bei der Ausführung von Apps muss die Trusted-Connector-Plattform sicherstellen, dass sich Apps nicht gegenseitig beeinträchtigen, die Ausführungsplattform selbst manipulieren oder Daten unkontrolliert preisgeben. Letzteres erfordert, dass Apps zunächst nicht in der Lage sein dürfen, ausgehende Verbindungen zu initiieren, sondern dieses Recht entweder explizit erteilt werden muss oder sämtliche Kommunikation über einen vertrauenswürdigen Referenzmonitor (\glqq Core Platform\grqq{}) erfolgt. Die Anwendungssicherheit des Trusted Connector umfasst damit die folgenden Kernaspekte:

\begin{itemize}
  \item Die Authentizität und Integrität von Apps ist sichergestellt.
  \item Apps können selbstständig keine Daten preisgeben, jede Kommunikation (in/out) wird kontrolliert und protokolliert.
  \item Apps laufen in Konnektoren strikt voneinander isoliert.
\end{itemize}

\end{multicols}

\chapter{Identitäts- \& Access-Management}

Das IIoT stellt besondere Anforderungen an das Identitäts- und
Access-Management (IAM), da im Gegensatz zu klassischen
Enterprise-Architekturen keine zentrale Instanz mehr über Zugriffe entscheiden
kann. Ziel des IAM, das durch den Trusted Connector implementiert wird, ist
daher, dem jeweiligen Konnektor-Betreiber die Kontrolle über seine Daten zu
geben und diese \emph{Autorisierung} von der darunterliegenden
\emph{Authentisierung} und \emph{Verschlüsselung} zu trennen. Authentisierung
und Verschlüsselung sind notwendige Voraussetzung für eine sichere
Kommunikation, die nicht durch Dritte abgehört oder modifiziert werden kann.
Ein Zugriff auf Daten erfordert jedoch zusätzlich eine Autorisierung, die von
verifizierten Identitätsattributen der Gegenstelle abhängt. Diese Attribute
können unterschiedlich stark beglaubigt sein und von einer unbestätigten
Selbstauskunft bis hin zu geprüften und zertifizierten Attributen reichen.

\vspace*{.5em}
\begin{multicols}{2}
\paragraph{Identitäten und Attribute}
Eine Identität im Industrial Data Space bezeichnet einen Trusted Connector und
seinen Betreiber. Sie besteht aus mehreren Komponenten, durch die eine
zunehmend starke Identifizierung realisiert wird. Zunächst werden Konnektoren
über ein \emph{X509v3-Zertifikat} identifiziert, das an ihren Hostnamen
gebunden ist. Für Konnektoren, die über das Internet erreichbar sind, können
diese Zertifikate durch einen öffentlichen ACME-Dienst erstellt werden, für
interne Konnektoren können unternehmensinterne ACME-Server oder selbst generierte und mit einer eigenen CA signierte 
Zertifikate verwendet werden. Auf dieser Stufe ist bereits eine vertrauliche
und authentische Kommunikation zwischen Konnektoren möglich, allerdings sind
noch keine weitergehenden Informationen über die Eigenschaften des Konnektors
oder seines Betreibers verfügbar. Solche Informationen werden in Form von
Identitätsattributen bereitgestellt, die durch einen Attribute
Provider verwaltet werden. 
Mittels der Identitätsattribute werden Angaben zum
Betreiber eines Konnektors, sowie zum Sicherheitsniveau des jeweiligen
Konnektors in einer Form hinterlegt, die es Datenanbietern ermöglicht
Zugriffskontrollrichtlinien auf Basis dieser Attribute festzulegen. So wird es
möglich, nur Konnektoren eines bestimmten Betreibers Zugriff auf Daten zu
erteilen, oder nur Konnektoren mit einem Mindestniveau an Sicherheit
zuzulassen.
Identitätsattribute des Attributeproviders sind von diesem beglaubigt. Nicht beglaubigte Identitätsattribute können vom Betreiber eines Konnektors selbst angegeben und werden von diesem in Form einer Selbstauskunft bereitgestellt. Ihre inhaltliche Richtigkeit ist also nicht sichergestellt und sie sind lediglich als eine ungeprüfte Beschreibung des Konnektors aus Sicht des (authentischen) Betreibers eines Konnektors zu verstehen.
Im Gegensatz dazu werden bestätigte Identitätsattribute durch eine unabhängige
Zertifizierungsstelle verifiziert. Hierzu definiert der Industrial Data Space
Zertifizierungskriterien und einen Prüfprozess, an dessen Ende die Erteilung
einer Zertifizierung und die Signatur der jeweiligen Identitätsattribute steht.

Auf diese Weise kann grundsätzlich jedermann die IDS-Infrastruktur verwenden, indem er einen Trusted Connector lediglich mit einem X509v3-Zertifikat betreibt, das bis auf die Richtigkeit des Hostnamens keine weiteren Zusicherungen gibt. Andererseits können Betreiber von einer höheren Sicherheitsstufe ihrer Konnektoren profitieren, in dem sie Zugang zu höherwertigen Diensten erlangen, die ihre Daten nur vertrauenswürdigen und ggf. zertifizierten Konnektoren bereitstellen. Die Voraussetzung hierfür ist, dass Datenanbieter und -konsument den selben Attribute Provider verwenden und diesem vertrauen. Dabei ist es jedoch nicht erforderlich, dass für die gesamte IDS-Infrastruktur nur ein Attribute Provider existiert. Vielmehr können mehrere Attribute Provider Identitätsattribute verschiedener Kontexte verwalten und parallel verwendet werden.
So ist es beispielsweise möglich, das Unternehmen in Konsortien und Projekten zusammenarbeiten und für den Zweck dieser Zusammenarbeit einen gemeinsamen Attribute Provider verwenden, bei dem sich alle Mitglieder des Konsortiums registrieren und fortan als solche ausgewiesen werden.

\begin{Figure}
  \centering
  \includegraphics[width=\textwidth]{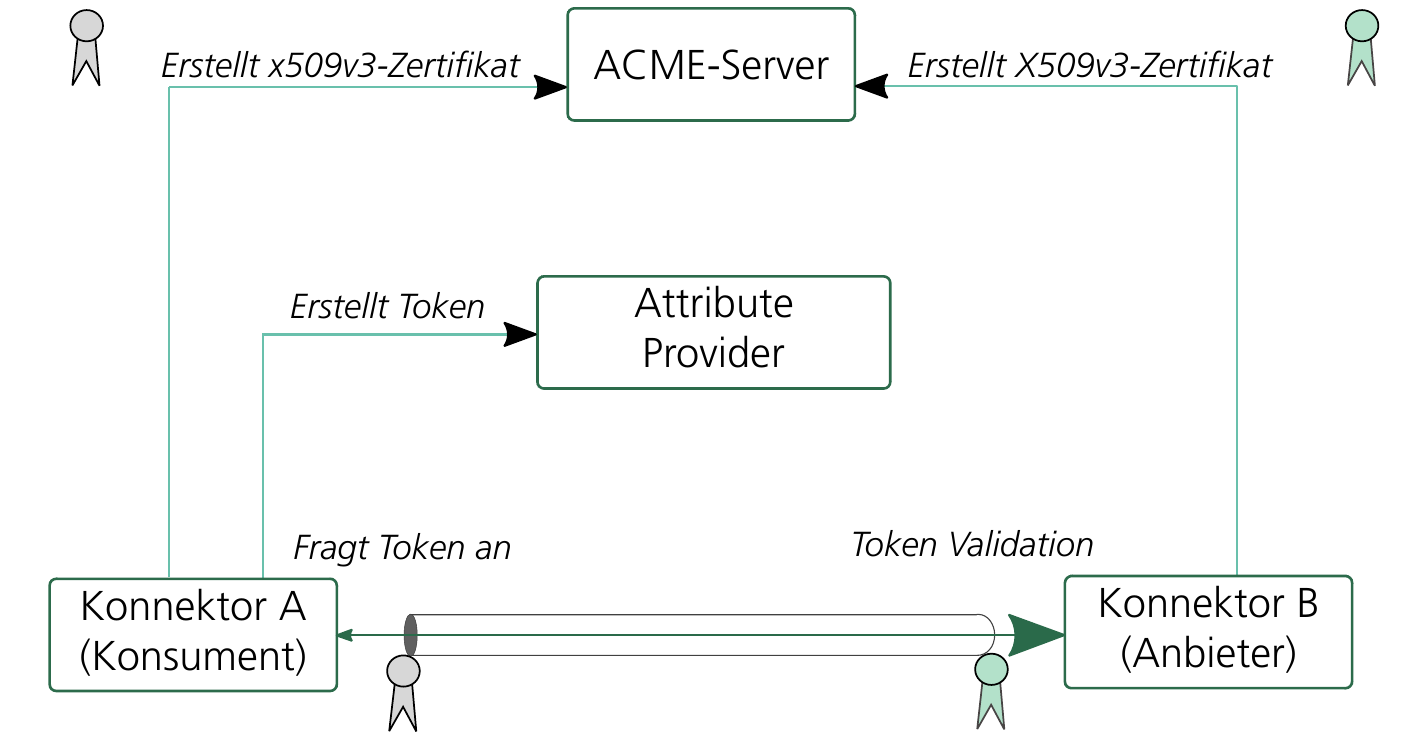}
  \captionof{figure}{Komponenten des Identitätsmanagement}
  \label{fig:idm}
\end{Figure}

\paragraph{Technische Umsetzung}

Für das Identitäts- und Access-Management verwendet der Trusted Connector Standardprotokolle aus dem Web-Umfeld. Beim erstmaligen Initialisieren eines Trusted Connector wird ein Public-/Private-Schlüsselpaar, sowie ein Signing Request für den Hostnamen des Konnektors erzeugt und mittels ACME-Challenge verifiziert. Für öffentlich erreichbare Konnektoren empfiehlt sich hierfür ein öffentlicher Dienst, dessen CA in vielen Clients als vertrauenswürdig betrachtet wird. Für interne Konnektoren kann entweder ein eigener ACME-Server zum Einsatz kommen, oder es werden selbstsignierte Zertifikate im Trusted Connector erstellt und manuell ausgetauscht. \\
Diese Zertifikate werden während der Verbindung zwischen Trusted Connectors über das IDS-Protokoll für den Aufbau der TLS-Verbindung benutzt -- sowohl für die Client-Authentisierung, als auch für die Verschlüsselung.
Ist die TLS-Verbindung etabliert, so werden im Rahmen des IDS-Protokolls JSON
Web Tokens (JWT) für den Zugriff auf Ressourcen des Trusted
Connectors etabliert. Hierzu stellt der Client (Datenkonsument) einen
OAuth2.0-konformen Token Request an den Attribute Provider und authentisiert
sich dabei mittels TLS Client Credentials. Über einen OAuth 2.0 \emph{Assertion
Flow} erhält er ein JWT mit bestätigten oder unbestätigten
Identitätsattributen, signiert vom Attribute Provider. Dieses sendet er an den
Datenanbieter, der auf Basis der enthaltenen Attribute wiederum entscheiden
kann, ob er den Zugriff gestattet oder ablehnt. Wurde die Verbindung auf diese
Weise etabliert, können fortan Daten über die bestehende Sitzung ausgetauscht
werden, ohne dass eine erneute Authentisierung erforderlich ist.
\end{multicols}

\chapter{Der Trusted Connector als sichere Ausführungsplattform}

Konnektoren sind die Edge-Gateways, über die interne Datenquellen in der IDS-Infrastruktur bereitgestellt und genutzt werden können. Sie sind das Kernelement der Sicherheitsarchitektur des IDS \cite{IDS2017} und realisieren sowohl die Protokolle für das Identitäts- \& Access-Management und zur sicheren Datenübertragung, als auch eine vertrauenswürdige Ausführungsumgebung für \emph{Apps} zur Datenvorverarbeitung und -Analyse.

\vspace*{1em}
\begin{multicols}{2}
Ein Trusted Connector verfügt über Sicherheitsmechanismen, mit denen die
Integrität des Software-Stacks, die Vertraulichkeit und Integrität der Daten sowie die Isolation von Apps sichergestellt
werden können. Zusammen ermöglichen diese Eigenschaften neue
Anwendungsszenarien, die mit bisherigen Sicherheitsarchitekturen nicht
realisiert werden können:

\paragraph{Höherwertige Datenangebote bei höherer Sicherheit} Konnektoren
können grundsätzlich in verschiedenen Sicherheitsstufen existieren -- als
einfaches Software-Programm auf einem herkömmlichen Rechner, bis hin zu
sicheren integrierten Hard- und Softwarestacks auf eingebetteten Systemen. Das
Sicherheitsniveau jedes Konnektors lässt sich mit Hilfe einer
Remote-Attestation feststellen und analog zu den Identitätsattributen
unterschiedlicher Stärke für die Entscheidung von Zugriffsanfragen verwenden.
So können Betreiber eines Trusted Connector festlegen, dass sie ausschließlich
mit Konnektoren mit einem nachweisbar integeren Software-Stack und
vertrauenswürdigen Apps kommunizieren möchten.

\paragraph{Datenverarbeitung an der Quelle} In vielen Fällen werden Rohdaten
von Sensoren für Analysezwecke benötigt. Im IIoT-Kontext sind diese Daten
jedoch hochkritisch, da sich aus ihnen detaillierte Informationen über interne
Produktionsabläufe gewinnen lassen. Eine Lösung ist es, die kritischen Daten
innerhalb des Unternehmens zu belassen und stattdessen die Analysedienste
in Form von \emph{Apps} zu den Daten zu transportieren. Hierzu muss allerdings
sichergestellt sein, dass Apps zuverlässig ausgeführt und nicht manipuliert
werden. Des Weiteren muss bei der Ausführung fremder Apps durch den
Daten-Provider garantiert werden, dass diese keinerlei Zugriff auf das
Netzwerk, andere Apps oder den Trusted Connector selbst erhalten, sondern von
anderen Prozessen isoliert bleiben und die zu verarbeitenden Daten lediglich
über eine definierte Schnittstelle erhalten.

\paragraph{Übersicht Trusted Connector-Architektur}

Der Trusted Connector ist ein Software-Stack, der im wesentlichen aus den
Komponenten Kernel, Container Management Layer (CML) und Core Container
besteht. Beim Kernel handelt es sich um einen Linux Kernel, der den Apps
eine POSIX-kompatible Laufzeitumgebung bereit stellt und zugleich alle
Kommunikation kontrolliert. Dies erfolgt insbesondere mit Linux-Namespaces,
Control Groups, Capabilities und einem Linux Security Module (LSM).
Optional können zudem auch die KVM-Kernel-Module verwendet werden.

Oberhalb des Kernels befindet sich das Container Management Layer (CML), das
den Lebenszyklus der Container, die Verifikation von Container-Images, sowie
das Nachladen von Containern aus dem App-Store übernimmt. Der Trusted Connector
unterstützt zwei CML-Implementierungen: Docker und trust-me. Die Vorteile des
Docker-Ökosystems sind seine weite Verbreitung, die Unterstützung
verschiedenster Plattformen -- angefangen bei eingebetteten Systemen wie dem
Raspberry Pi bis hin zu skalierenden Cloud- und Cluster-Plattformen wie Amazon
AWS/ECS und Kubernetes. Diese Mächtigkeit kann jedoch auch von Nachteil sein,
wenn dedizierte eingebettete Geräte für sicherheitskritische Anwendungen
benötigt werden. Hier kommt das trust-me CML zum Zug, dessen Vorteile im
Fokus auf starke Container-Isolation sowie nativer Unterstützung für
Sicherheitsmechanismen wie Secure Boot, eine TPM-gestützte Full Disk Encryption
(FDE) und in einer netfilter-Separierung von Containern bestehen.

Container beinhalten Apps, die Funktionalität zur Verarbeitung und Analyse von
Daten bereitstellen können. Apps sind grundsätzlich isoliert und haben keinen
Zugriff auf Speicher oder Dateisystem anderer Apps oder den CML. Auch wird
ihnen kein Netzwerkinterface zugeteilt, über das sie mit dem externen Netz oder
anderen Apps kommunizieren könnten. Ihre Berechtigungen können weiter durch
Kernel-Capabilities, sowie durch das Linux Security Module kontrolliert werden.
Die Zuteilung von Systemressourcen wie CPU und Speicher lässt sich ebenfalls
Container-spezifisch festlegen.
In diesem Zustand sind Apps zwar vor dem Zugriff aufeinander und auf das darunterliegender System geschützt, aber wenig nutzbringend. Um sie nutzen zu können, müssen sie mit Daten versorgt werden und ihre Ergebnisse abrufbar sein können.
Dies geschieht durch einen besonderen, privilegierten Container -- die sogenannte \emph{Core Platform}. Die Core Platform ist der einzige Container, der ein Netzwerkinterface mit externem Zugriff erhält und mit jedem einzelnen App-Container über eine Netzwerk-Bridge kommunizieren kann. Somit muss jede Kommunikation innerhalb eines Trusted Connector über die Core Platform laufen. Innerhalb der Core Platform befinden sich Mechanismen zur Datenflusskontrolle, das IDS-Protokoll, über das Verbindungen mit externen Trusted Connectors aufgebaut werden können, sowie die Management-Schnittstellen, über die die Einstellungen des CML verwaltet werden können.

\begin{Figure}
  \centering
  \includegraphics[width=.9\textwidth]{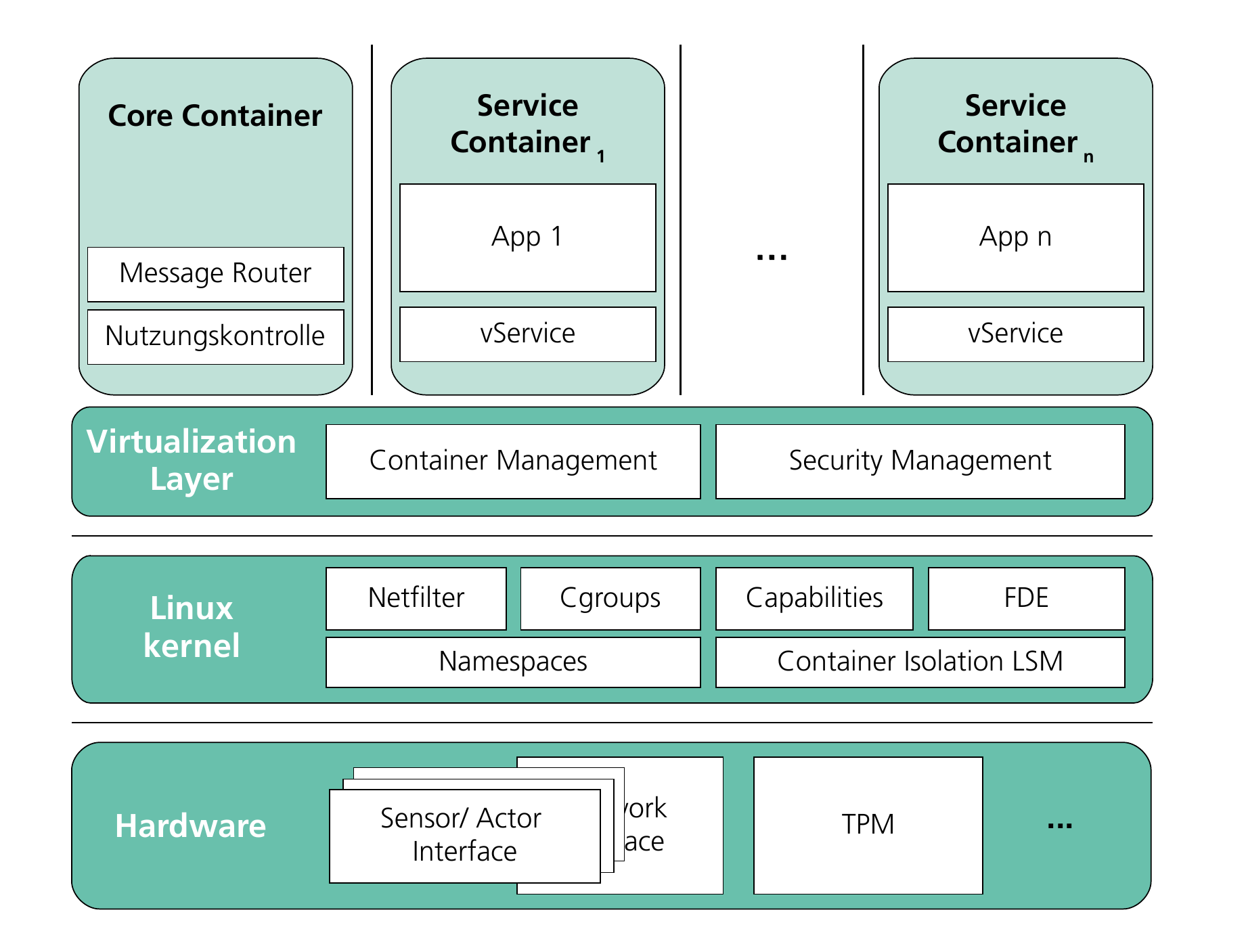}
  \captionof{figure}{Komponenten des Trusted Connector-Stack}
  \label{fig:konnektorarchitektur}
\end{Figure}

\paragraph{Integrität der Software durch Remote Attestation} Die Integrität des
Software-Stacks eines Trusted Connector wird durch Hardwarevertrauensanker in
Form eines Trusted Platform Module (TPM) realisiert. Ein TPM ist ein
dediziertes vertrauenswürdiges Hard- oder Softwaremodul, das eine sichere Verwaltung
kryptografischer Schlüssel, Verschlüsselungs- und Signaturoperationen, sowie
einen sicheren Speicher bereitstellt. 
Mittels eines von der Trusted Computing
Group standardisierten \emph{Secure Boot}-Prozesses verifiziert der Trusted
Connector beim Start des Systems die Integrität aller Betriebssystemkomponten
und hinterlegt ihre Hashwerte in den sog. PCR-Registern.
Konnektoren verwenden diese Werte während des Verbindungsaufbaus, um sich
gegenseitig im Rahmen des IDS-Protokolls mittels einer Remote-Attestation die
Integrität ihres Software-Stacks nachzuweisen. Die Remote-Attestation
unterstützt hierbei drei Sicherheitslevel: Bei Level 0 wird de facto keine
Attestierung durchgeführt -- die Konnektoren tauschen lediglich den Hinweis
aus, dass sie über kein TPM oder keinen Secure-Boot-Prozess verfügen. Level 1
umfasst einen Integritätsnachweis von Boot-Firmware, Bootloader, Kernel und
Core Platform. Dies sind damit die unveränderlichen Bestandteile des
Trusted Connector, die nicht modifiziert werden können, ohne die PCR-Hashes zu
verändern. Da natürlich Software-Updates dieser Komponenten nach wie vor
möglich sein müssen, werden TPM 2.0 custom policies verwendet, um die
Entscheidung über die Korrektheit eines PCR-Wertes an eine externe
\emph{Software Authority} zu delegieren. \\
Eine Remote-Attestation auf Level 2 umfasst zusätzlich zu den Komponenten des Level 1 auch die Messung der installierten Apps. Hierzu werden die Container-Images der Apps zusammen mit ihren Meta-Daten, d.h. der jeweiligen App-Beschreibung geprüft und die Integrität der dabei entstehenden PCR-Werte zusammen mit dem \emph{measurement log} wiederum durch die Software Authority bestätigt.

\paragraph{Core Platform}

Sämtliche Kommunikation zwischen Apps und zwischen Trusted Connectors erfolgt über die Core Platform. Diese dient daher als zentraler Punkt für das Erstellen von Audit-Logs und zur Kontrolle von Datenflüssen zwischen Apps. Der Vorteil hierbei ist, dass Apps nicht vertrauenswürdig sein müssen und nicht von IDS-spezifischen Schnittstellen abhängen. Anstatt Apps spezifisch für den IDS zu entwickeln, können Konnektor-Betreiber so existierende Apps (z.B. Docker-Container) in den Trusted Connector laden und in ihre Datenfluss-Konfiguration einbinden. So lange die App eines der Protokolle verwendet, die vom Message Router der Core-Plattform unterstützt werden, kann sie in eine Datenverarbeitungskette in Form einer \emph{Message-Route} eingebunden werden. Message-Routes orchestrieren Nachrichten zwischen einzelnen Apps, konvertieren sie in die erforderlichen Formate und stellen sie schließlich über das IDS-Protokoll anderen Konnektoren zur Verfügung, wobei die Core Platform jeden einzelnen Verarbeitungsschritt im Rahmen der Datenflusskontrolle prüft. Neben der zentralen Kommunikationschnittstelle beinhaltet die Core-Platform auch die Schnittstellen zum Management des Trusted Connectors. Betreiber konfigurieren ihren Konnektor entweder über ein Webinterface oder über eine textbasierte Konsole.
\end{multicols}

\chapter{Datennutzungskontrolle mit dem Trusted Connector}

Souveränität über Daten zu wahren, ist das oberste Ziel des Industrial Data Space und zwingende Voraussetzung für Anwendungsfälle, die über das reine Teilen von ohnehin öffentlichen Daten hinausgehen. Mit dem Trusted Connector als vertrauenswürdigem Hardware-/Software-Stack verfügen die IDS-Teilnehmer über einen Endpunkt, über den sich fortgeschrittene Techniken der Datennutzungskontrolle umsetzen lassen. Datennutzungskontrolle unterscheidet sich von herkömmlicher Zugriffskontrolle dadurch, dass nicht der Zugriff auf eine Ressource (z.B. einen Dienst) zu einem Zeitpunkt kontrolliert wird, sondern die Nutzung der Daten über die Zeit hinweg. Dies ist erforderlich, um typische Anforderung an die Kontrolle der Datennutzung zu realisieren, wie die folgenden Beispiele zeigen.

\begin{multicols}{2}
\let\footnote\parnote
\paragraph{Beispiel: Einhaltung von Datenschutzanforderungen}

Bei der Verarbeitung persönlicher Daten müssen Konnektor-Betreiber jederzeit Auskunft über den Verbleib dieser Daten geben können, sie auf Anfrage des Eigentümers löschen und sicherstellen, dass die Verarbeitung ausschließlich dem angegebenen Zweck folgt. Was schon Betreiber herkömmlicher Anwendungen mit zentralen Datenbanken vor Herausforderungen stellt, wird für Datenanbieter im Industrial Internet of Things zur Mammutaufgabe. Die Herkunft jedes einzelnen Datums muss nachverfolgbar sein und während der Verarbeitung muss beurteilt werden können, in welchem Verarbeitungsschritt Daten als personenbezogen gelten.

\paragraph{Beispiel: Nutzungsrestriktionen für bereitgestellte Daten}

In dem Moment, in dem Daten von einem Dienst abgerufen werden, verliert dieser typischerweise die Kontrolle über sie. Für viele IIoT-Anwendungen ist es jedoch erforderlich, dass auch sensible Daten ausgetauscht und mit Nutzungsrestriktionen versehen werden können. Solche Nutzungsrestriktionen umfassen beispielsweise die Auflage, die Daten nach einer bestimmten Zeit zu löschen, eine Einschränkung des Einsatzzwecks oder die Forderung, Daten nicht weiterzuleiten.

\paragraph{Datenflusskontrolle mit LUCON}
Der Trusted Connector verwendet standardmäßig das LUCON Policy-Framework zur Kontrolle von Datenflüssen zwischen Apps und Konnektoren. LUCON markiert Daten mit \emph{Labels}, sobald sie den Trusted Connector erreichen. Im weiteren Verlauf der Datenverarbeitung in einer Message-Route werden Labels über Apps hinweg transportiert und durch diese ggf. erweitert, modifiziert, oder entfernt. Abhängig von den Labels, die an einer Nachricht haften, können mit LUCON Einschränkungen und Auflagen -- sogenannte \emph{Obligations} -- verbunden werden. So kann beispielsweise verhindert werden, dass als \emph{privat} markierte Daten an externe Dienste versendet werden oder zunächst einen Anonymisierungsdienst durchlaufen müssen. Durch Obligations können Aktionen auf dem Trusted Connector ausgeführt werden oder zusätzliche Auflagen in Form von \emph{sticky policies} mit den Daten verknüpft werden. Aktionen auf dem Trusted Connector umfassen beispielsweise das Logging von Nachrichten für Audit-Zwecke oder das Löschen von Daten aus einem Dienst. Sticky Policies sind Richtlinien, die zusammen mit den Daten im Rahmen des IDS-Protokolls an den Trusted Connector der Gegenstelle übermittelt werden. Dies können einerseits wiederum LUCON-Policies sein, mit denen die Gegenstelle die Datenflussanforderungen des Datenanbieters umsetzen wird, oder aber Nutzungsrestriktionen in Form von ODRL. Die Open Digital Rights Language (ODRL) ist ein Standard zur Spezifikation von Nutzungseinschränkungen für Daten und kann beispielsweise dazu verwendet werden, die Nutzungsdauer, -häufigkeit oder die Art der zulässigen Operationen auf Daten zu definieren. Im Gegensatz zu LUCON-Policies können ODRL-Anforderungen jedoch nicht automatisch durchgesetzt werden, sondern dienen als nachweisbare Vereinbarung zwischen Konnektor-Betreibern.

\paragraph{Auditierbare Datenverarbeitung}
Neben der aktiven Durchsetzung von Datenflussanforderungen lässt sich mit LUCON auch überprüfen und ggf. nachweisen, dass die Datenverarbeitung im Trusted Connector die Anforderungen einhält. Betreiber können so auf einen Blick erkennen, ob die konfigurierten Message-Routen jederzeit ihren Anforderungen entsprechen, oder ob sie unter bestimmten Konstellationen unzulässige Datenflüsse bewirken könnten (die wiederum zur Laufzeit blockiert würden). Hierzu verwendet LUCON intern eine formale Repräsentation von Message-Routen und Richtlinien und prüft mittels Model-Checking mögliche Verletzungen der Richtlinien durch die Route.

\begin{Figure}
  \centering
  \fboxsep=0pt
  \fboxrule=0.2pt
  \fcolorbox{gray}{lightgray}{\includegraphics[width=.9\linewidth]{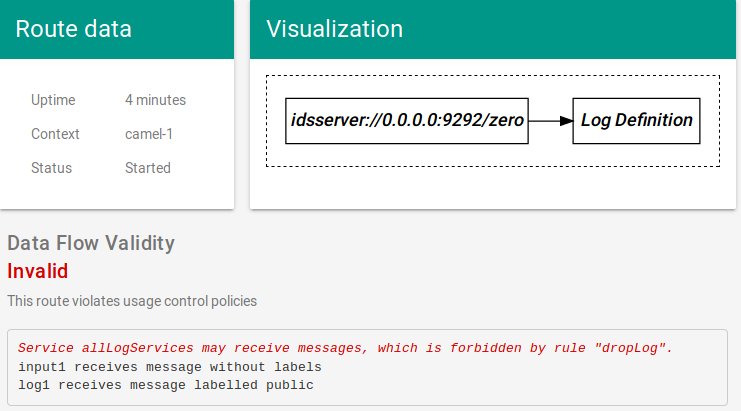}}
  \captionof{figure}{Warnung bei Message-Routen, die Nutzungsrestriktionen verletzen}
  \label{fig:lucon-validation}
\end{Figure}

\paragraph{Technische Umsetzung}

Der Trusted Connector verwendet Apache Camel als Message-Router -- eine quelloffene Lösung zur Realisierung von Enterprise Integration Patterns, die in vielen großvolumigen Produktiv-Anwendungen erprobt wurde. Für die Erstellung von LUCON-Policies existiert ein Xtext\footnote{\url{https://www.eclipse.org/Xtext/}}-basierter Editor für die Eclipse-Plattform, der den Entwickler mit Auto-Vervollständigung und Syntax-Highlighting unterstützt und Policies automatisch kompiliert. Die kompilierten Policies werden anschließend über die Administrationsschnittstelle in den Trusted Connector geladen und dort ohne weiteres Zutun des Benutzers zur Verifikation der Camel-Routen, sowie zur Durchsetzung der Richtlinien verwendet.

\begin{Figure}
  \fcolorbox{gray}{lightgray}{\includegraphics[width=.97\linewidth]{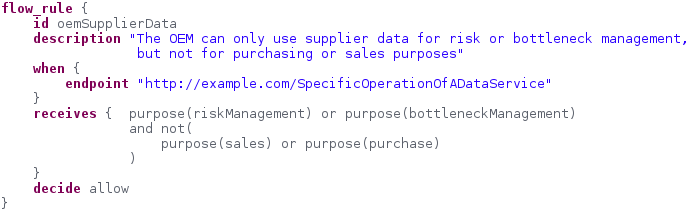}}
  \captionof{figure}{LUCON Datenfluss-Richtlinie in Eclipse-Editor}
  \label{fig:lucon-example}
\end{Figure}

Obligations können in die Core Platform geladen werden und stehen dann
für die Durchsetzung von Richtlinien zur Verfügung. So wird bspw. die
Anforderung von ODRL-Nutzungsrestriktionen in Form einer Obligation realisiert,
die ODRL-Daten an die jeweilige Nachricht bindet und an den Trusted Connector
der Gegenstelle übermittelt.

\begin{lstlisting}[caption={ODRL-Nutzungsrestriktion},captionpos=b]
<http://example.com/policy:1111> a odrl:Offer ;
  odrl:permission [
      odrl:action odrl:use ;
      odrl:target <http://example.com/SpecificOperationOfADataService> ;
      odrl:assigner <http://example.com/Supplier> ;
      odrl:assignee <http://example.com/OEM> ;
      odrl:constraint [
          odrl:purpose ids:RiskManagement;
          odrl:purpose ids:BottleNeckManagement;
      ];
  ];
  odrl:prohibition [
      odrl:action odrl:use ;
      odrl:target <http://example.com/SpecificOperationOfADataService> ;
      odrl:assigner <http://example.com/Supplier> ;
      odrl:assignee <http://example.com/OEM> ;
      odrl:constraint [
          odrl:purpose ids:Purchasing;
          odrl:purpose ids:Sales;
      ];
  ].
\end{lstlisting}

Die Kombination aus nachweisbarer Vertrauenswürdigkeit der Trusted Connector-Plattform und Datennutzungskontrolle erlaubt es Betreibern, auch sensible Daten anzubieten und dabei ihren rechtlichen Anforderungen technisch nachweisbar nachzukommen.
\parnotes
\end{multicols}

\newpage

\pagestyle{empty}
\begin{textblock*}{29mm}(0cm,0cm)%
  \begin{tcolorbox}[colframe=fraunhofergray,colback=fraunhofergray,sharp corners,halign=flush left,valign=center,height=5.5cm,width=21cm]%
  \hspace*{1cm}\Huge{Kontakt}
  \end{tcolorbox}%
\end{textblock*}%

\begin{textblock*}{6cm}(2cm,10cm)%

\textbf{\color{fraunhoferblue} Ansprechpartner}

Julian Schütte\\
Gerd Brost\\

Fraunhofer AISEC\\
Parkring 4, 85748 Garching bei München\\

\Telefon\ \href{tel:+49893229986-292 }{+49 (0) 89 3229986-292}\\
\Email\ \href{mailto:info@aisec.fraunhofer.de}{info@aisec.fraunhofer.de}
\end{textblock*}%

\begin{textblock*}{6cm}(13cm,10cm)%

\textbf{\color{fraunhoferblue} Autoren}

Dr. Julian Schütte\\
\emph{Service \& Application Security}

Gerd Brost\\
\emph{Service \& Application Security}

Sascha Wessel\\
\emph{Secure Operating Systems}
\end{textblock*}%

\begin{textblock*}{10cm}(2cm,17cm)%
\textbf{\color{fraunhoferblue} Open-Source-Projekt}

\setlength{\tabcolsep}{-6pt}
\renewcommand{\arraystretch}{0}
\begin{tabular}{m{2.8cm} m{8cm}}
\includegraphics[width=1.8cm]{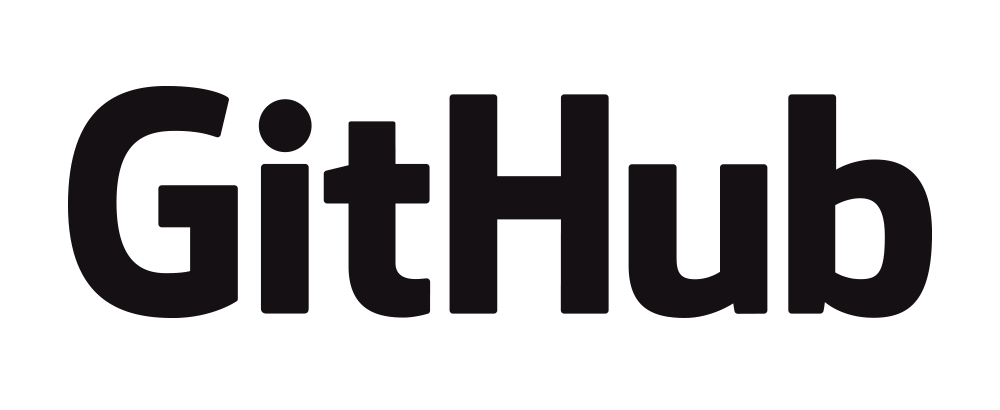} & \vspace*{5pt} https://github.com/industrial-data-space
\end{tabular}
\end{textblock*}%

\begin{textblock*}{16cm}(2cm,19.5cm)%
\textbf{\color{fraunhoferblue} Referenzen}
\vspace{-3.75cm}
\onecolumn{
\bibliographystyle{ieeetr}
\bibliography{references}}

\end{textblock*}%

\begin{textblock*}{16cm}(2cm,24cm)%
\textbf{\color{fraunhoferblue} Förderrahmen}

Der Trusted Connector ist eine Komponente des Industrial Data Space, gefördert durch das Bundesministerium für Bildung und Forschung im Rahmen des Projektes InDaSpacePlus (Förderkennzeichen 01|S17031). Anwendungen des Trusted Connector werden im Rahmen der Aktivitäten des Forschungsclusters Cognitive Internet Technologies CIT gefördert.
\end{textblock*}%

\afterpage{\blankpage}
\clearpage
\newpage

%

%
%

\end{document}